# Origins of Large Voltage Hysteresis in High Energy-Density Metal Fluoride Lithium-Ion Battery Conversion Electrodes

Linsen Li[1¶], Ryan Jacobs[2], Peng Gao[3†], Liyang Gan[1], Feng Wang[3], Dane Morgan[2*], and Song Jin[1*]

[1] Department of Chemistry, University of Wisconsin-Madison, Madison, WI 53705, USA.

[2] Department of Materials Science and Engineering, University of Wisconsin-Madison, Madison, WI 53706, USA.

[3] Sustainable Energy Technology Division, Brookhaven National Laboratory, Upton, NY 11973, USA.

[¶] Present address: Department of Materials Science and Engineering, Massachusetts Institute of Technology, Cambridge, MA 02139, USA

† Present address: School of Physics, Peking University, Beijing 100871, China.

*Corresponding author e-mail: ddmorgan@wisc.edu and jin@chem.wisc.edu

**ABSTRACT**

Metal fluoride and oxides can store multiple lithium-ions through conversion chemistry to enable high energy-density lithium-ion batteries. However, their practical applications have been hindered by an unusually large voltage hysteresis between charge and discharge voltage-profiles and the consequent low energy efficiency (< 80%). The physical origins of such hysteresis are rarely studied and poorly understood. Here we employ *in situ* X-ray absorption spectroscopy (XAS), transmission electron microscopy (TEM), density-functional-theory (DFT) calculations, and galvanostatic intermittent titration technique (GITT) to first correlate the voltage profile of iron fluoride (FeF$_3$), a representative conversion electrode material, with evolution and spatial distribution of intermediate phases in the electrode. The results reveal that, contrary to conventional belief, the phase evolution in the electrode is symmetrical during discharge and charge. However, the spatial evolution of the electrochemically active phases, which is controlled by reaction kinetics, is different. We further propose that the voltage hysteresis in the FeF$_3$ electrode is kinetic in nature. It is the result of Ohmic voltage drop, reaction overpotential, and different spatial distributions of electrochemically-active phases (i.e. compositional inhomogeneity). Therefore, the large hysteresis can be expected to be mitigated by rational design and optimization of material microstructure and electrode architecture to improve the energy efficiency of lithium-ion batteries based on conversion chemistry.

**INTRODUCTION**

Lithium-ion battery (LIB) technology has revolutionized portable electronics and been considered a promising solution for future large-scale energy applications.[1, 2] Current LIBs function through intercalation chemistry, which consists of topotactic insertion/removal of Li$^+$ into/from the host lattice of electrode materials, for example LiCoO$_2$, LiFePO$_4$, LiNi$_{1/3}$Mn$_{1/3}$Co$_{1/3}$O$_2$, and graphite.[3] Despite good rate capability and long cycle-life, the energy density achievable with intercalation (500−600 Wh kg$^{-1}_{\text{active material}}$) is inherently limited by the number of interstitial sites in the host lattice (typically < 1 per formula).[1] New battery materials and/or new chemistries with higher specific energy densities are clearly desirable.[1, 3-5]

The discovery of reversible multiple-lithium storage in metal fluorides/oxides in the early 2000s opens up promising opportunities for high energy-density storage that does not necessarily depend on available interstitial sites.[6-11] Instead, it is realized through a heterogeneous conversion reaction ($MF_x + xLi^+ + xe^- = M + xLiF$ or $M_xO_y + 2yLi^+ + 2ye^- = xM + yLi_2O$). The last decade has witnessed tremendous advances in preparation of nanostructured conversion electrode materials that exhibit high specific energy densities (and power capability).[4] However, their practical applications have been deterred by an unusually large voltage hysteresis (voltage gap) between discharge and charge steps. Similar hysteresis has also been observed in other high energy-density battery chemistries such as $Li-O_2$ and Li-sulfur (S). This hysteresis of conversion electrode materials range from several hundred mV to ~2 V,[4] comparable to that of a $Li-O_2$ battery[5] but much higher than that of a Li-S battery (200–300 mV)[5] or a typical intercalation electrode material (several tens mV)[12] at similar rates. It leads to a high degree of round-trip energy inefficiency (< 80% energy efficiency) that is unacceptable in practical applications. To overcome this challenge, it is necessary to understand its physical origins, which have remained elusive, as most work has focused on material synthesis and electrochemical testing[4].

Iron fluorides ($FeF_3$ and $FeF_2$) are among the most studied conversion electrode materials due to their very high specific energy densities (> 1000 Wh $kg^{-1}$) and better reversibility as compared with other fluorides (such as $CuF_2$), especially after nanostructuring and/or mixing with conductive carbon.[9-11, 13-26] Notably, $FeF_3$ is one of the very few conversion electrode materials[27, 28] whose voltage hysteresis has been previously studied, by either density functional theory calculations within the generalized gradient approximation (DFT-GGA)[14] or electrochemical measurements.[19, 20] The current belief is that a large portion of the voltage hysteresis originates from the asymmetric reaction pathways during discharge and charge and consequently the existence of different intermediate phases results in the large split in electrochemical potential.[14, 19, 20] According to the DFT-GGA calculations and associated models, FeF$_3$ is first lithiated to Li$_x$[Fe$^{3+}_{1-x}$Fe$^{2+}_x$]F$_3$ before full reduction of Fe to LiF/Fe during discharge, and a series of Fe$^{3+}$-containing compounds (Li$_{3-3x}$Fe$^{3+}_x$F$_3$) form sequentially during charge before formation of a defect trirutile FeF$_3$.[14] However, the understanding of the reaction pathways is not fully justified by *ex situ* solid nuclear-magnetic resonance (NMR),[15] pair distribution function analysis (PDF),[15] transmission electron microscopy (TEM),[13] *in situ* X-ray absorption spectroscopy,[25] and *in situ* X-ray spectroimaging experiments,[26] all of which suggest the existence of Fe$^{2+}$-containing intermediate phases during and/or after charge. Additionally, different interpretations of the intermediate phases that form during discharge exist between the *ex situ*[15] and *in situ* experimental works[25] (e.g., trirutile Li$_{0.5}$FeF$_3$ vs. rhombohedral Li$_{0.92}$FeF$_3$, respectively). These discrepancies in reaction pathways have constrained the understanding of voltage hysteresis and therefore need to be reconciled through systematic and definitive studies into both thermodynamic and kinetic aspects of the reaction mechanism.

Here we use *in situ* synchrotron X-ray absorption spectroscopy (XAS) to track changes in Fe oxidation states and local bonding structure during cycling of three iron fluoride model samples, FeF$_2$ nanowires (NWs), FeF$_3$ NWs, and FeF$_3$ microwires (MWs). Combining results from *in situ* TEM experiments and hybrid functional DFT calculations (HSE06), we show that the reaction pathway is symmetrical and as follows: rhombohedral FeF$_3$ → trirutile Li$_{0.25}$FeF$_3$ → trirutile Li$_{0.5}$FeF$_3$ ⇌ rutile FeF$_2$ + LiF ⇌ Fe + 3LiF. However, reaction homogeneity (completeness and spatial evolution of each electrochemical reaction) is strongly influenced by reaction kinetics in these sequential multiple-step reaction processes. Based on the new mechanistic understanding and results from galvanostatic intermittent titration technique (GITT) experiments, we show that the large voltage hysteresis of the $FeF_3$ electrode is due to iR (Ohmic) drop, reaction overpotential, and difference in apparent potentials which are result of different spatial distributions of electrochemically active phases. These results have general implications for understanding voltage hysteresis in other conversion electrode materials and provide the basis of new strategies to minimize its adverse effect.

**RESULTS**

**Iron fluoride model samples**

We prepared $FeF_3$ NWs[18] and MWs[26] according to previous work and synthesized the $FeF_2$ NWs for the first time via thermal reduction of $FeF_3$ NWs using a small amount of glucose at 450 °C under flowing argon (see Methods for synthetic details; see morphology in Figure S1a-c). The phase identities of these samples are confirmed using powder X-ray diffraction (PXRD, Figure S1d). Further TEM characterization reveals that all of these wire samples are polycrystalline and made of attached particle domains (Figure S2). We chose these materials as the model samples because of their higher electrochemical activity at room temperature compared with other iron fluoride samples.[11, 15, 16] They can all reach near theoretical capacity at a moderate current rate (Figure S3), which is critical for finishing the *in situ* experiments in a reasonable amount of beam time and collecting useful mechanistic information.

***In Situ* XAS on an $FeF_2$ electrode**

We first studied the reaction mechanism of an $FeF_2$ electrode using *in situ* XAS as a comparison to the $FeF_3$ electrode, because Li-$FeF_2$ is a simpler conversion system (ideally only $Fe^{2+}$ and $Fe^0$ are involved) than Li-$FeF_3$ ($Fe^{3+}$, $Fe^{2+}$, and $Fe^0$ are involved). Figure 1a shows the electrochemical profile of a Li/$FeF_2$ battery discharged at a current of C/12 (1 C = 571 mA g$^{-1}$ for FeF$_2$) to ~1.2 V. The discharge cut-off voltage was chosen based on previous literature[15] and to ensure that the FeF$_2$ electrode achieved near theoretical capacity (2 Li per FeF$_2$). During charge, the battery was charged at a rate of C/6 to 4.2 V (the current was doubled due to limited time). After the constant-current charging step, a constant-voltage charging step was applied at 4.2 V until the current dropped to ~C/50. Fe $K$-edge XAS spectra were collected every 18 min during the electrochemical cycling so that the change in average states of lithiation ($x_{Li}$) was +0.05 per spectrum during discharge and −0.1 per spectrum during charge (see $x_{Li}$ for each spectrum in Table S1). The change in Fe oxidation state and local bonding structure were monitored, respectively, by X-ray absorption near-edge structure (XANES, Figure 1b) and extended X-ray absorption fine-structure spectroscopy (EXAFS, Figure 1c).

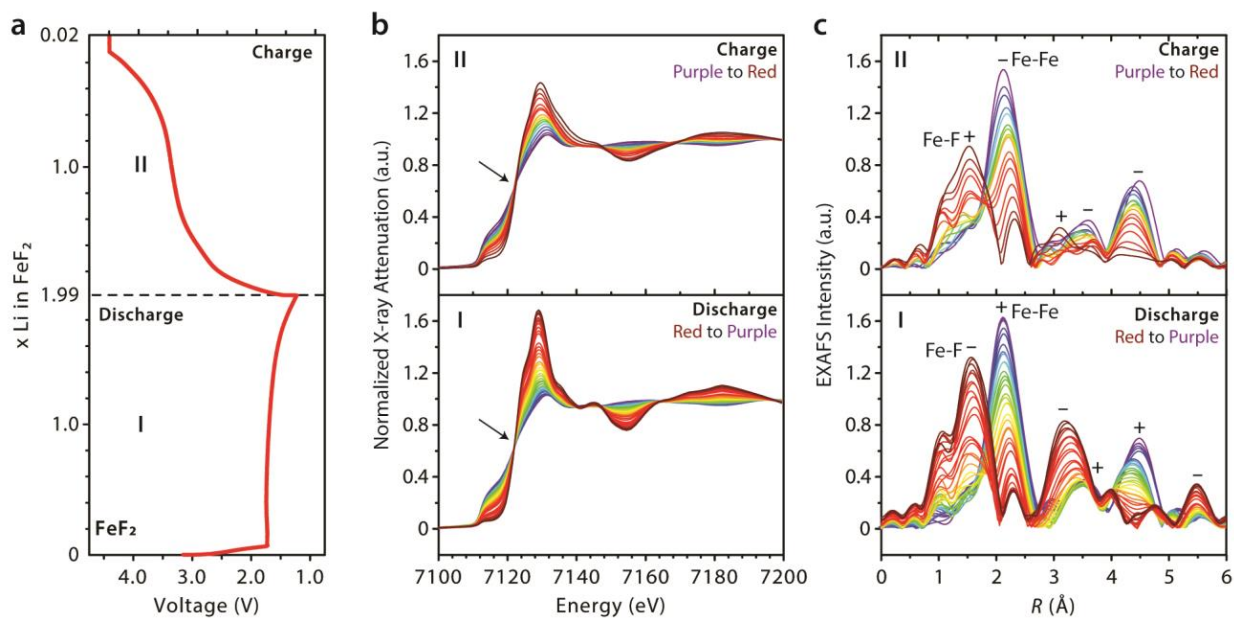

**Figure 1.** *In situ* **XAS results on an FeF$_2$ electrode. a,** Voltage profile of an FeF$_2$ NW electrode discharged at a current rate of 1/12 C (1 C = 571 mA g$^{-1}$ for FeF$_2$) and recharged at a current rate 1/6 C. **b,** XANES and **c,** EXAFS spectra taken at every 18 min during active discharge (+0.05 $x_{Li}$

per spectrum) and charge (−0.10 $x_{Li}$ per spectrum), respectively. The black arrows indicate the isosbestic points shared by the XANES spectra.

During discharge ($x_{Li}$ = 0→1.99 Li per FeF$_2$, region **I** in Figure 1), we observed that the absorption edge of the XANES spectra gradually shifted toward lower energies, the white line intensity concurrently decreased, and an isosbestic point is shared by all the spectra (indicated by the arrow), indicating a two-phase conversion reaction FeF$_2$ + 2Li$^+$ + 2e$^-$ → Fe + 2LiF. Accordingly, in the EXAFS patterns, the intensity of FeF$_2$-related peaks gradually decreases (marked with "−") as the intensity of Fe-related peaks increases (marked with "+"). Standard EXAFS patterns of FeF$_2$ and Fe are shown in Figure S4 for comparison. It appears that FeF$_2$ became lithium-saturated and started to decompose into LiF and Fe quickly, as we observed the metallic Fe-related EXAFS peaks as soon as the cell voltage hit the plateau at ~1.75 V (<0.05 Li insertion). This result is consistent with the calculations done by local environment dependent GGA+$U$ (DFT-LD-GGA+$U$)[29] but different from those by DFT-GGA,[14, 17] which show that lithiation of rutile FeF$_2$ first produces a mixture of metallic Fe and sub-stoichiometric Li$_{0.5}$FeF$_3$ prior to producing LiF.[14, 17]

Interestingly, we also observed progressive shifts in the EXAFS peaks during discharge. For example, the Fe-related peak shifted towards smaller $R$ values as the discharge reaction proceeded, decreasing from ~2.3 Å (~ 0 Li insertion) to ~2.1 Å (~1.99 Li insertion). Further EXAFS fittings revealed that the Fe-Fe bond length gradually decreases (Figure S5), which indicates that the Fe nanoparticles formed initially have a larger lattice constant than those formed later during discharge (conversion), consistent with results from the previous *in situ* TEM experiments on $FeF_2$ nanoparticles.[17]

During charge ($x_{Li}$ = 1.99→0.02 Li per $FeF_2$, region **II** in Figure 1), the changes in both XANES spectra and EXAFS patterns closely mirror what occurred during discharge, indicating that Fe and LiF are gradually reconverted into a rutile phase highly similar to $FeF_2$ in the local structure. This rutile phase most likely nucleates first at the interface between the electrolyte and the active particles (now made of LiF/Fe nanocomposites), where $Li^+$ ions are extracted and transferred into the electrolyte most easily. When the cell was charged to > ~3.3 V, some Fe was over-charged to +3 oxidation state, as evidenced by the absorption edge of the XANES spectra shifting toward higher energy and the Fe-F related peaks in the EXAFS patterns shifting towards smaller $R$ value than that of the pristine $Fe^{2+}F_2$ electrode. We also performed linear combination fitting analysis (LCA; see fitting examples in Figure S6a and fitting parameters in Table S1) to estimate the relative mole fraction of different Fe oxidations states (Figure S7). The results reveal that the electrode was charged back to a multiple phase mixture containing Fe, $Fe^{2+}$, and $Fe^{3+}$, not the pure rutile $FeF_2$ phase.

### *In Situ* XAS and TEM on $FeF_3$ electrodes

Next, we studied the reaction mechanism of an $FeF_3$ NW electrode. An Li/$FeF_3$ battery was discharged at a current of C/10 (1 C = 712 mA $g^{-1}$) to 1.0 V and then charged at a rate of C/10 to 4.5 V, after which a constant-voltage charging step was applied at 4.5 V until the current dropped to ~C/50 (Figure 2a). Fe $K$-edge XAS spectra were collected every 18 min during the electrochemical cycling so that the change in average states of lithiation ($x_{Li}$) was about +0.09 per spectrum during discharge and −0.09 per spectrum during charge (see $x_{Li}$ for each spectrum in Table S2). The electrochemical profile (Figure 2a) is divided into four different regions based on features observed in the XANES spectra (Figure 2b) and EXAFS patterns (Figure 2c).

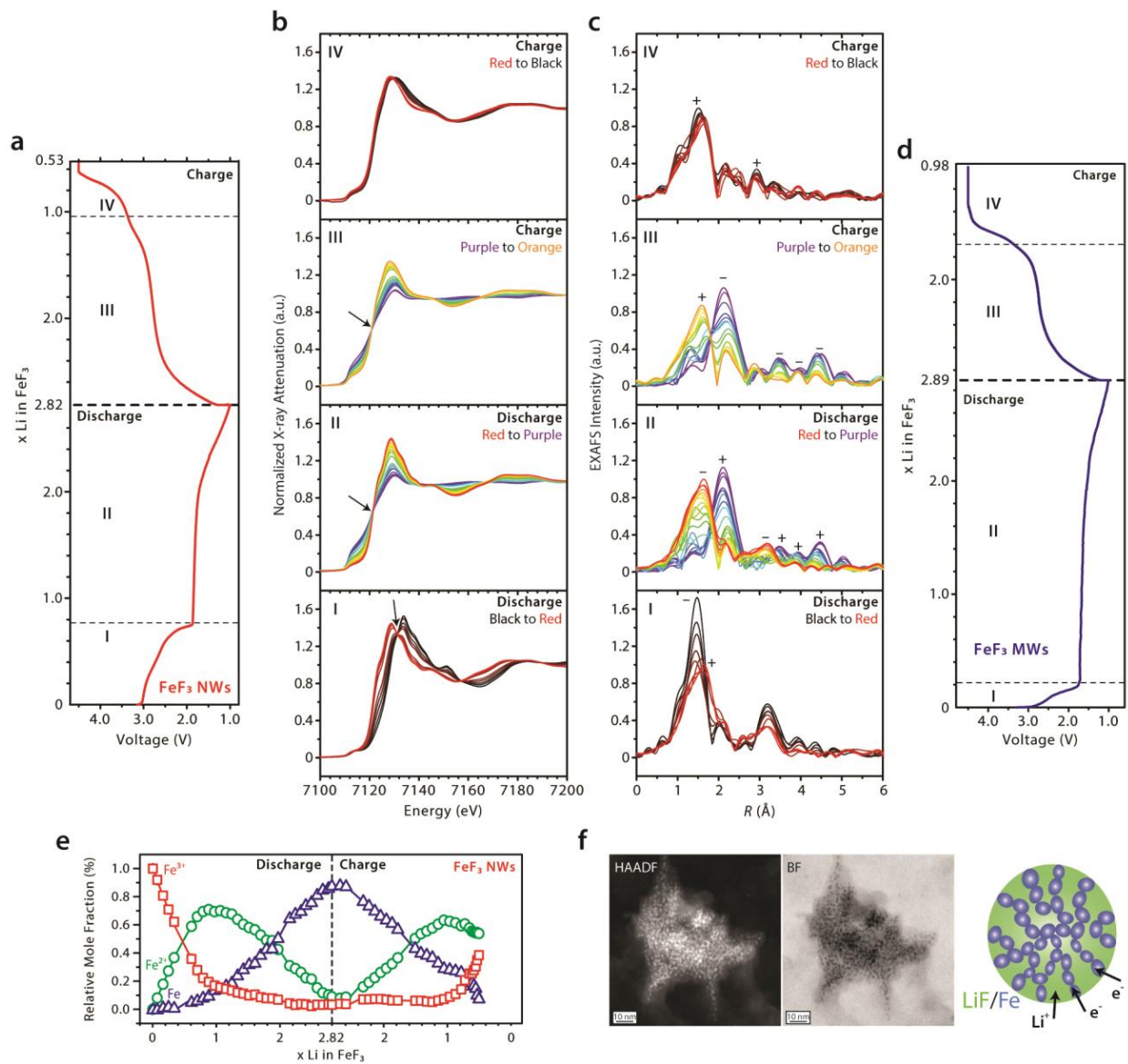

**Figure 2. *In situ* XAS and TEM on FeF$_3$ NW electrodes. a,** Voltage profile of an FeF$_3$ NW electrode discharged and recharged at a current rate of 1/10 C (1 C = 712 mAh g$^{-1}$). **b,** XANES and **c,** EXAFS spectra taken every 18 min during active discharge (+0.09 $x_{Li}$ per spectrum) and charge (−0.09 $x_{Li}$ per spectrum). The black arrows indicate the isosbestic points. **d.** Voltage

profile of an FeF$_3$ MW electrode discharged and recharged at a current rate of 1/10 C, shown as a comparison to the NW electrode. **e.** Phase evolution during the cycling of the FeF$_3$ NW electrodes, which is estimated by linear combinational fitting analysis of the XANES spectra. **f.** High-angle annular dark-field, bright-field STEM images and schematic illustration showing the microstructure of a bundle of fully lithiated FeF$_3$ NWs, which is made of interconnected Fe domains surrounded by LiF. The STEM images were recorded in the *in situ* TEM experiment.

In discharge region **I** in Figure 2a-c ($x_{Li}$ = 0→0.78 Li per FeF$_3$), the Fe$^{3+}$ in FeF$_3$ is gradually reduced to Fe$^{2+}$ with Li uptake, as evidenced by the shift in absorption edge of the XANES spectra. Meanwhile, the change in EXAFS peak position and intensity indicates that the local structure deviates from that of the original rhombohedral FeF$_3$ and becomes increasingly rutile-like, which resembles rutile FeF$_2$ after 0.78 Li insertion (Figure S8). We further carried out EXAFS fitting and found out that the first set of data collected after discharge region **I** ($x_{Li}$ = 0.79) could be best modeled using the scattering paths generated from rutile FeF$_2$ (see fitting results in Figure S9 and Table S2). For the FeF$_3$ electrode at a state of lithiation of $x_{Li}$ = 0.79, the coordination numbers are slightly smaller while coordination distances are slightly larger than those of rutile FeF$_2$ standard, indicative of a slightly defective rutile structure. In another Li-FeF$_3$ NW battery discharged at a slower rate of 1/20 C, we observed two different isosbestic points in the XANES spectra (Figure S10) in region **I**, which may be indicative of two different trirutile Li$_x$FeF$_3$ phases. No additional isosbestic points were observed afterwards. In the subsequent discharge region **II** in Figure 2a-c ($x_{Li}$ = 0.78→2.82 Li per FeF$_3$), in the changes of the absorption edge, white line intensity, and EXAFS peak position and intensity, we observed changes that are highly similar to those which occurred during the discharge of the FeF$_2$ NW electrode (Figure 1, region **I**), indicating reduction of a rutile $Fe^{2+}$-containing phase to metallic Fe. We also studied the phase and microstructural evolution of the $FeF_3$ NW electrode by *in situ* TEM electron diffraction (ED) and scanning transmission electron microscopy (STEM), which show results consistent with the *in situ* XAS experiments. The *in situ* ED patterns (Figure S11 and Movie S1) show that the $FeF_3$ NWs were first lithiated to form rutile $FeF_2$ phase before full reduction to metallic Fe. The *in situ* STEM (Movie S2) shows that the lithiation reaction initiated from the surface and propagated toward the core of each electrochemically active domain. Notably, after being fully lithiated, the microstructure of the $FeF_3$ NW electrode is similar to that of the fully lithiated $FeF_2$ nanoparticles.[16, 17] Nanocomposites consisting of bicontinuous LiF/Fe networks were formed (Figure 2f). The average size of the Fe domains is 2–3 nm. As the starting point for the charge process, this microstructure also provides the key to understanding the phase evolution during delithiation. The charge reaction (delithiation) most likely first initiates from the surface of the active particles (now made of LiF/Fe nanocomposites), where $Li^+$ ions are extracted and transferred into the electrolyte most easily.

We performed LCA (See fitting examples in Figure S6b and fitting parameters in Table S3) to estimate the relative mole fraction of different Fe oxidation states (Figure 2e) during the discharge (and charge) of the $FeF_3$ electrode. We found that a noticeable amount of Fe (> 5%) already existed at $x_{Li} = 0.61$ and $Fe^{3+}$, $Fe^{2+}$, and Fe coexisted in the electrode between $x_{Li} = 0.61$ to 2.15 (<5 % $Fe^{3+}$, Figure 2e, Table S3). This result indicates that the reduction of $Fe^{2+}$ to Fe had already started on the outside ($Fe^{2+}$-containing region) surface before the first reduction ($Fe^{3+}$ to $Fe^{2+}$) fully finished in the whole electrochemically active particle. An interesting question is what Li composition in the Li-$FeF_3$ system is necessary for metallic Fe to begin forming, since different values $x_{Li} = 0.75$, 0.92, or 1.0 were previously reported in $FeF_3$/carbon nanocomposite samples.[14, 15, 25] In order to understand this issue, we studied the reaction mechanism of FeF$_3$ MWs for comparison (Figure S12), which consist of larger particle domains than the NWs (Figure S2). The FeF$_3$ MW/Li cell was cycled at the same rate as the FeF$_3$ NW/Li cell, but its voltage profile (Figure 2d) shows different features. The first sloping plateau at higher voltages (region **I**) is much shorter but the second flat plateau (region **II**) is much longer than those of the FeF$_3$ NW/Li cell. Accordingly, the XANES spectra and EXAFS patterns of the FeF$_3$ MW electrode (Figure S12a-c) is also different from those of the NW electrode (Figure 2a-c). LCA fittings performed on the XANES spectra (fitting examples in Figure S6c and fitting parameters in Table S4) reveal that a noticeable amount of metallic Fe started to exist at $x_{Li} = 0.31$ (> 5% mole fraction) and $Fe^{3+}$, $Fe^{2+}$, and Fe coexisted in the electrode until $x_{Li} = 2.41$ ($Fe^{3+} < 5\%$, Figure S13 and Table S4). We also note that the energy density of the Li-FeF$_3$ MW battery is lower than the Li-FeF$_3$ NW battery due to the loss of the high voltage plateau.

The comparison between the reaction behavior of FeF$_3$ NWs and MWs reveals how reaction homogeneity and voltage profiles (energy density) are affected by the size of the active domains, which correlates with the reaction kinetics. As shown by the *in situ* STEM experiment (Movie S2), the lithiation reaction of FeF$_3$ proceeds from the surface to the core of each electrochemically active domain. In FeF$_3$ MWs that consist of larger active domains (and thus smaller surface area) than the NWs, the applied current (1/10 C) is more likely to exceed what the reaction kinetics (Li$^+$ and/or electron transport) can keep up with. Therefore, the $Fe^{2+}$-containing rutile phase produced in the initial reduction (lithiation) on the outside is further lithiated to produce LiF and Fe early at $x_{Li} = 0.31$ before the interior FeF$_3$ domains can begin to react (see schematic illustration in Figure S14). The occurrence of the reduction of $Fe^{2+}$ to Fe dictates the voltage profile despite the presence of unreacted FeF$_3$ at the interior so that the second flat plateau becomes much longer (Figure 2d). By contrast, in the NWs, faster reaction kinetics (shorter distance for Li$^+$ and/or electron transport) allow the first reduction step (Fe$^{3+}$→Fe$^{2+}$) to further complete before Fe formation (at $x_{Li}$ = 0.61). Furthermore, in the FeF$_3$/carbon nanocomposite samples reported previously,[14, 15, 25] even smaller particle size, better electrical contact afforded by the carbon matrix, and small current densities likely facilitate the first reduction to complete even more, which explains the initial formation of Fe approaching $x_{Li}$ at 1.0. We note that $x_{Li}$ is better considered a measure of the state of lithiation averaged within the entire electrode and it does not necessarily reflect the stoichiometric information of the Li$_x$FeF$_3$ phase that readily extrudes Fe upon further lithiation, especially in the kinetically limited situations (such as the MWs). It is possible that in all cases Fe starts to form at $y$ = 1, but the existence of unreacted FeF$_3$ phase at the core of the active particles leads to $x_{Li}$ < 1. These results illustrate the critical role of reaction kinetics and inhomogeneity in governing the conversion processes and voltage curves for the FeF$_3$ conversion electrode and have implications for other electrode materials operating through sequential multiple-step processes.

Additionally, we compared the electrochemical capacity with the capacity estimated from the LCA fittings for the FeF$_2$ and FeF$_3$ NW electrode and found reasonable matches (Figure S15). These results suggest that electrolyte decomposition (or any other non-metal-center side reactions) does not contribute significantly to the observed discharge capacity in the iron fluoride conversion electrodes studied in this work (when low cut-off voltage ≥ 1 V is used). This reaction behavior is different from metal oxide conversion electrodes that are discharged to lower voltages (< 1 V),[28, 30-33] in which additional capacity was often observed.

During charge of the FeF$_3$ NW electrode (region **III** of Figure 2a-c, $x_{Li}$ = 2.82→1.03 Li per FeF$_3$), the changes in XANES spectra and EXAFS patterns not only mirror what occurred in discharge (region **II** of Figure 2a-c), but also highly similar to those observed during the charge of the $FeF_2$ electrode (region **II** of Figure 1). EXAFS fitting was also performed and the data could be best modeled using scattering paths generated from rutile $FeF_2$ and Fe (Figure S16). These results provide clear evidence that a rutile-$FeF_2$-like phase is formed during charge of the $FeF_3$ electrode. These findings are actually consistent with previous results from *ex situ* NMR and PDF experiments[15] but in disagreement with the DFT-GGA based reaction mechanism[14, 17], which suggested formation of $Fe^{3+}$-containing intermediate phases rather than the $FeF_2$ intermediate during charge. In fact, we only observed oxidation of $Fe^{2+}$ into $Fe^{3+}$ when the cell voltage exceeded ~3.3 V in charge region **IV** (Figure 2a-c, $x_{Li}$ = 1.03→0.53 Li per $FeF_3$), as evidenced by the absorption edge further shifting toward higher energies. Further, the local structure at the final state (0.53 Li per $FeF_3$) still resembles rutile $FeF_2$. According to EXAFS patterns, the Fe-F peak position is slightly smaller than that in rutile $FeF_2$ but larger than that in rhombohedral $FeF_3$ (Figure 2c) in $R$ value. Therefore, we suggest that some trirutile $Li_{0.5}FeF_3$ may exist at the end of the charge process.

**DFT calculations and reaction pathway of $FeF_3$ electrodes**

To corroborate the experimental findings, we performed a detailed multicomponent phase analysis using DFT calculations of materials in the Li-Fe-F ternary system. The DFT calculations were performed using GGA, GGA+$U$, and hybrid HSE functionals[34], which have been shown to more accurately reproduce experimental formation energies and Li insertion voltages for transition metal-containing compounds than GGA.[35, 36] The Li-Fe-F phase diagrams calculated using GGA and GGA+$U$ are shown in Figure S17 and S18, respectively. The results are consistent with those previously reported (GGA,[14, 17] GGA+$U^{29}$). The HSE phase diagram shown in Figure 3a is very similar to the GGA+$U$ diagram (Figure S18), with the exception that Li$_{0.25}$FeF$_3$ is not stable from GGA+$U$. As HSE is a somewhat more general method than GGA+$U$ (due to GGA+$U$ generally requiring a fitted $U$ for every transition metal), we chose to include and discuss our HSE results in the main text, and provide our GGA+U (and GGA) results in the Supporting Information section for comparative purposes. In all the phase diagrams presented in this work (Figure 3a, S17, and S18), the red dots represent stable phases, the black dots represent materials that were predicted to be unstable, and the purple dots are important composition points where no lithiated FeF$_2$ or FeF$_3$ materials were calculated due to an insufficient number of interstitial sites for Li insertion.

When examining the stable FeF$_2$ lithiation path (green dotted line in Figure 3a), FeF$_2$ immediately begins to dissociate upon lithiation to precipitate metallic Fe and LiF. This three-phase region persists over the entire lithiation path until $x_{Li}$ = 2, at which point the reduction from Fe$^{2+}$ to metallic Fe is complete and produces a two phase mixture of metallic Fe and LiF. This lithiation path is consistent with the *in situ* XAS results on the FeF$_2$ electrode (Figure 1, region **I**) and the previous DFT-LD-GGA+$U$ calculations[29] but clearly different the one predicted by DFT-GGA calculations.[14, 17] In delithiation of the Fe/LiF (1:2 in mole ratio), FeF$_2$ should be formed as the stable phase but Li$_{0.5}$FeF$_3$ may also be produced from FeF$_2$ if there is excess LiF, which is likely the case at the surface of the active particles and indeed observed by the *in situ* XAS experiment (Figure 1, region **II**).

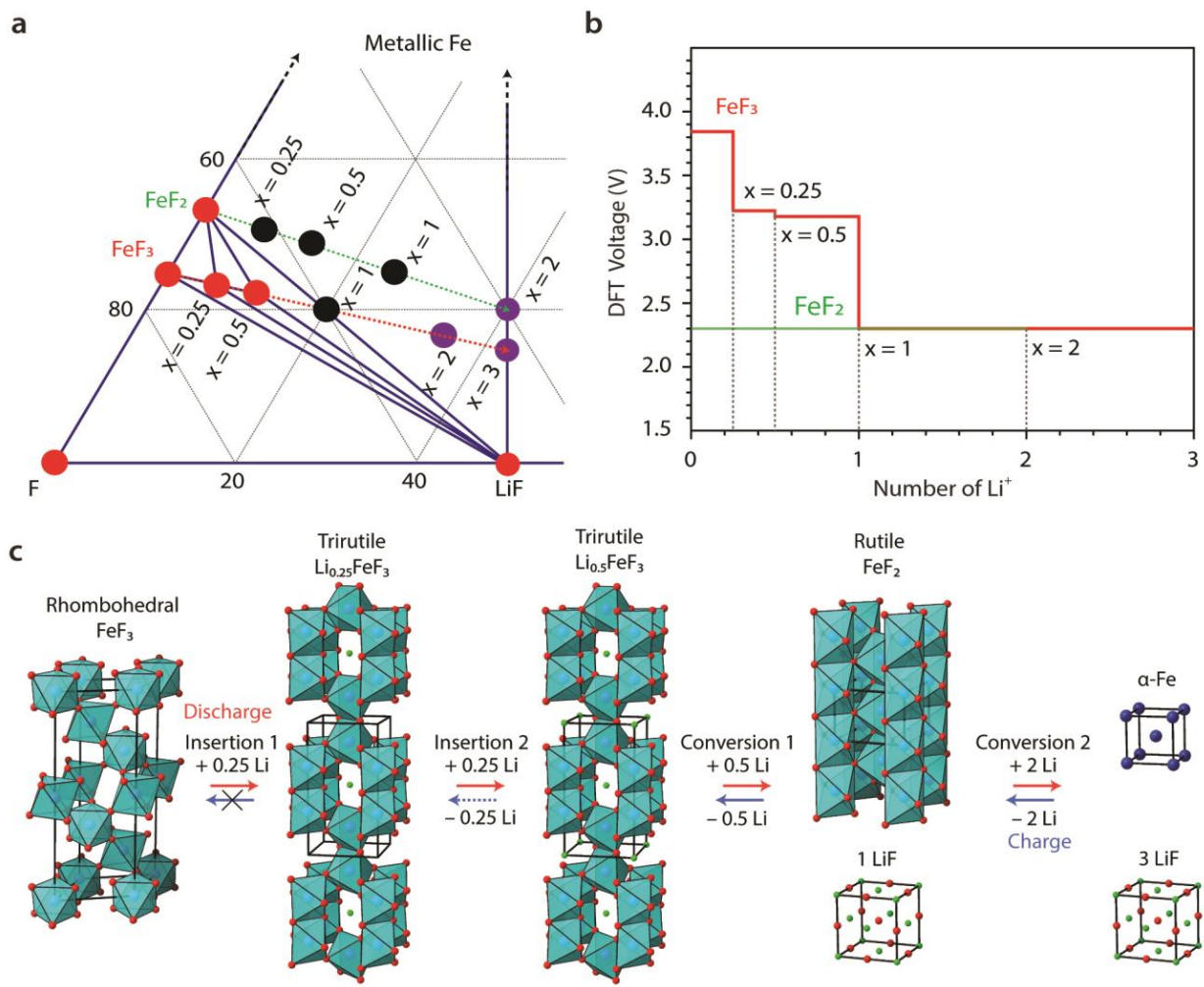

**Figure 3. DFT-HSE calculation results and FeF$_3$ reaction pathway. a,** DFT calculated Fe-Li-F phase diagram using the HSE approach. The lithiation pathways for FeF$_3$ and FeF$_2$ are indicated by the red and green dashed arrows, respectively. Red dots represent stable phases, black dots represent unstable lithiated phases and purple dots indicate potentially active compositions where no lithiated compound was calculated. The fraction of lithiation *x* for Li$_x$FeF$_2$ and Li$_x$FeF$_3$ are labeled for both pathways. **b,** Calculated DFT-voltage curves for FeF$_3$ and FeF$_2$ at different states of lithiation. **c,** Discharge and charge reaction pathways of the FeF$_3$ electrode and their crystal structures, which are derived from both the experimental and DFT

calculation results. Li, Fe, and F atoms are represented by green, blue, and red spheres. Ref. 14 was used as a guide for the range of Li compositions to test in these structures.

The stable lithiation path for $FeF_3$ (red dotted line in Figure 3a) shows direct Li intercalation when $x \leq 0.5$. Upon lithiation to $x_{Li} = 0.25$ ($Li_{0.25}FeF_3$), the pristine rhombohedral $FeF_3$ phase is no longer stable, and a phase change to the defected trirutile structure occurs. This defected trirutile phase is stable up to $x_{Li} = 0.5$ ($Li_{0.5}FeF_3$), after which dissociation to $FeF_2$ and LiF occurs because no more interstitial sites for lithium insertion are available. When $x_{Li} = 1$, all $Fe^{3+}$ has been reduced to $Fe^{2+}$ and the system is a two-phase mixture of $FeF_2$ and LiF. Further lithiation promotes the reduction of $Fe^{2+}$ to metallic Fe, which is exactly the same process as the lithiation of $FeF_2$. In delithiation of Fe/LiF (1:3 in mole ratio), $FeF_2$ should be formed as the stable phase first; $Li_{0.5}FeF_3$ can be produced later from the $FeF_2$ and the remaining LiF. These DFT-HSE calculation results are in good agreement with the *in situ* XAS results (Figure 2a-c) and the corresponding discussion on the $FeF_3$ electrode and the previous DFT-LD-GGA+$U$ calculations.[29]

Figure 3c shows the DFT lithiation voltages. Since they are representative of equilibrium voltages, in which case no polarization or overpotential is included, they are higher than those experimentally observed when a current was applied (Figure 1a and 2a). When the experimental battery is allowed to relax to approach equilibrium conditions, its voltages should become closer to the DFT calculated values. This trend is indeed seen from the GITT measurements after relaxation, which is discussed in more detail below.

Combining the results from the *in situ* XAS, TEM, and HSE-DFT calculations, we can now propose complete and consistent reaction pathways for $FeF_3$ (and $FeF_2$) electrodes (Figure 3c), which is quite symmetrical just like what the XAS data displays (Figure 1b, c and Figure 2b, c; better seen in surface contour plots of XANES and EXAFS in Figure S19). We note that kinetic limitations can cause one reaction not proceed completely over the entire particle domain before the subsequent one being forced to initiate in the pre-reacted region under galvanostatic condition, causing compositional inhomogeneity and less symmetrical phase evolution profile ($FeF_3$ MWs vs NWs, Figure S13 vs Figure 2e). This new proposed understanding is clearly different from the one proposed previously based on DFT-GGA calculations,[14, 17] which was the basis for understanding the large voltage hysteresis in $FeF_3$ and other conversion electrode materials. The previous model assumes that the electrochemical reaction is controlled by the slow diffusion of Fe so that Fe is oxidized to the highest oxidation state ($Fe^{3+}$) during charge in order to maximize lithium extraction. A series of $Fe^{3+}$-containing phases, such as spinel $Li_{15/8}Fe^{3+}_{3/8}F_3$, ilmenite $Li_{3/2}Fe^{3+}_{1/2}F_3$, and rutile $Li_{3/4}Fe^{3+}_{3/4}F_3$ are predicted to form sequentially during charge, which constituents a fundamentally different reaction pathway from that taken during discharge (reduction of rutile $Fe^{2+}F_2$ like phase to Fe).[14, 17] This model provided a seemingly reasonable explanation for the voltage hysteresis because the presence of different phases (and with Fe at different oxidation states) during discharge and charge would indeed lead to different potentials. However, our new mechanistic understanding clearly suggests that other mechanisms are responsible for the voltage hysteresis.

**GITT analysis**

To better understand the possible causes of the hysteresis, we performed GITT experiments on the $FeF_3$ NW and MW electrodes (Figure 4a). The cells were allowed to relax for 4 h after every 1 h discharging/charging at 50 mA g$^{-1}$. The GITT profiles are also divided into four regions based on the understanding of phase evolution in the FeF$_3$ NW and MW electrodes. Figure 4b provides a close-up view of the GITT process. In the discharge half-cycle (red curve), as soon as the current is removed, the voltage first suddenly increases a small amount, and then gradually increases as the electrode approaches equilibrium condition.[37] The opposite occurs in the charge half-cycle (blue curve in Figure 4b). We found that the voltage after the 4 h relaxation (V$_{relax}$, black dashed lines in Figure 4a) correlates with the composition of the electrodes inferred from the *in situ* XAS results. For example, since the Fe$^{3+}$F$_3$ phase in the MW electrode is reacted more slowly than that in the NW electrode during discharge, V$_{relax}$ observed in the MW electrode is higher initially (black dashed lines in Figure 4a). As the Fe$^{3+}$F$_3$ phase is consumed, the two V$_{relax}$ curves of the MW and NW electrodes become more comparable. Figure 4c shows how much the voltage relaxes after 4 h for the FeF$_3$ NW and MW electrodes, respectively, at different states of lithiation. During discharge, the voltage relaxes a lot more in the MW electrode than the NW electrode (Figure 4c), which is a direct consequence of inhomogeneity: the intermediate phase Fe$^{2+}$F$_2$ is already being further lithiated to produce LiF and Fe on the outside despite the presence of unreacted FeF$_3$ at the interior of the active particle. After all FeF$_3$ is consumed, the magnitude of the voltage relaxation in the MW and the NW electrode becomes comparable. Similar analysis based on reaction homogeneity can be made for the charging process. Figure 4d shows the remaining voltage difference at the same state of lithiation between charge and discharge steps after the 4 h relaxation (V$_{gap}$) for both the MW and NW electrodes. V$_{gap}$ can become slightly smaller based on its changing trend if the relaxation time is further increased. However, it did not become zero after 24 h relaxation in a separate experiment. In addition, It was previously reported by Liu et al that a 280 mV voltage-gap remained even after 72 h relaxation (measured at states of lithiation of $x_{Li}$ = ~2.0)[19].

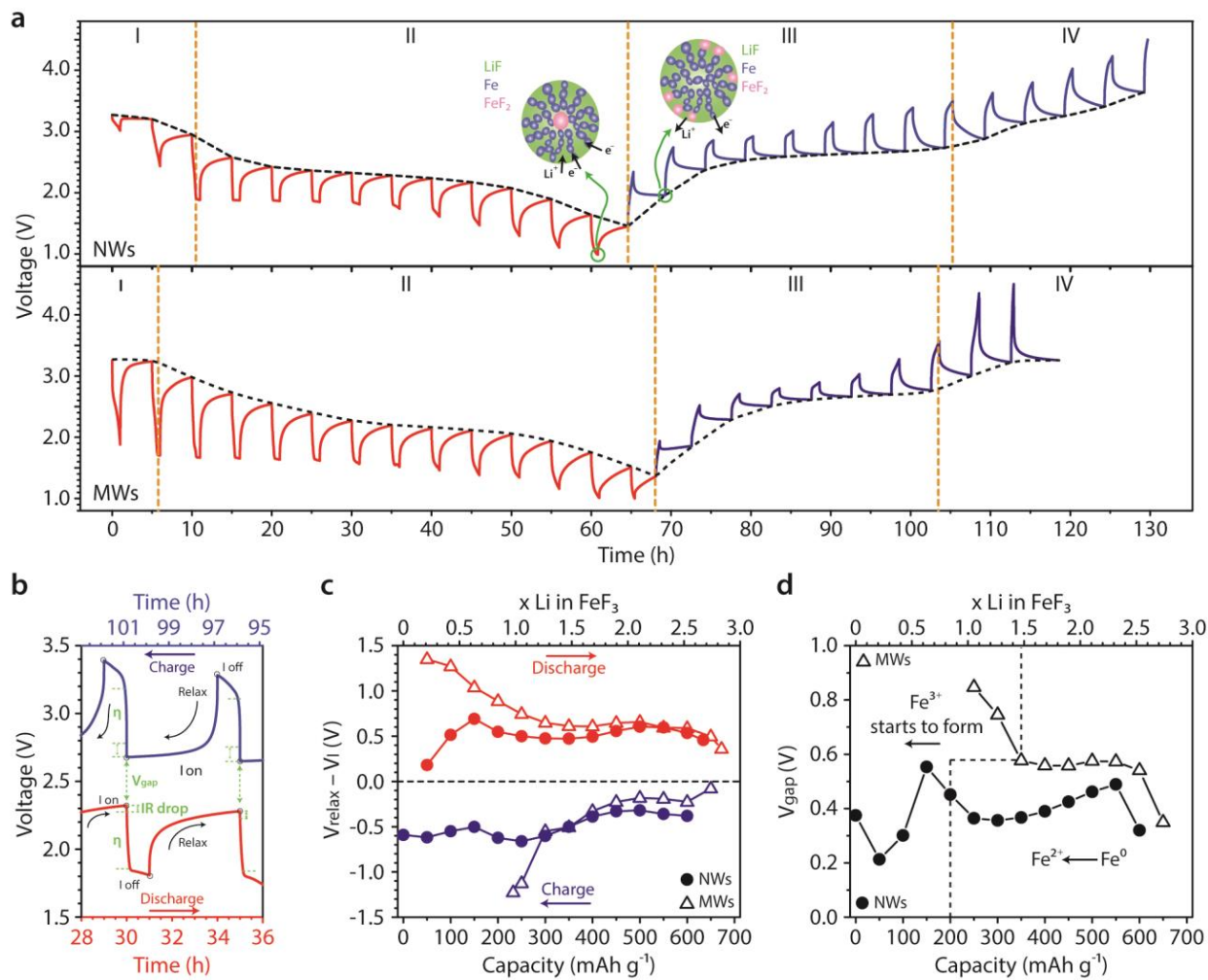

**Figure 4. GITT of FeF$_3$ NW and MW electrodes. a,** GITT profiles of an FeF$_3$ NW electrode and a MW electrode. The cells were allowed to relax for 4 h after every 1 h discharging/charging at 50 mA g$^{-1}$. Inset is a schematic illustration of the microstructures of an active domain in the FeF$_3$ electrode at states close to full lithiation (x$_{Li}$ = ~3), which are drawn based on the STEM results. **b,** Close-up view of the GITT curve for the NW electrode. IR drop, reaction overpotential ($\eta$), and the remaining voltage difference after relaxation (V$_{gap}$) are marked to show the components that contribute to the large voltage gap during cycling. **c,** Voltage change after the 4 h relaxation at different states of discharge and charge of the NW and MW electrodes, respectively. **d,** Voltage difference ($V_{gap}$) between charge and discharge steps after the 4 h relaxation at the same state of lithiation of the NW and MW electrodes, respectively.

**DISCUSSION**

**Proposed Origins of voltage hysteresis in FeF$_3$ conversion electrodes**

By integrating all experimental and theoretical simulation results, we can identify the following components from the GITT that contribute to the voltage hysteresis observed at non-zero current (see Figure 4b). The first one is the iR voltage drop, which is the sudden voltage jump after the current is removed and typically < 100 mV in our measurements. The second component is the reaction overpotential ($\eta$) that is required to nucleate and grow new phases, drive mass transport, and overcome the interfacial penalty for making nanophases. This overpotential is manifested in the voltage plummet when the current is applied and the spike when the current is removed. However, its magnitude is not straightforward to quantify using the GITT results, because the active particles undergo phase transformations and cannot achieve a truly homogeneous composition over the entire particle simply through Li$^+$ diffusion during the relaxation. Reverse-step potentiostatic intermittent titration technique (PITT) may be a more suitable approach to provide the quantitative evaluation, according to which the overpotential is 300 mV for the conversion reaction (reduction of intermediate product FeF$_2$ to LiF/Fe, measured at $x_{Li}$ = 1.2 per FeF$_3$) and 70 mV for reconversion reaction (LiF/Fe to FeF$_2$, measured at $x_{Li}$ = 1.2 per FeF$_3$).[20]

The third component that leads to the hysteresis but has not been considered in detail in previous literature, is the difference in spatial distribution of electrochemically active phases during discharge and charge as well as the way these phases are connected in the electrochemical system (i.e. access to $Li^+$ and electron). For example, we can infer from the *in situ* TEM and XAS results that at states of lithiation close to $x_{Li} = 3$ per $FeF_3$, during discharge the intermediate phase $FeF_2$ is located at the interior of the active particles while Fe/LiF are on the outside and have contact with the electrolyte and current collector (see schematic illustration in Figure 4a inset, left); in contrast, during charge the intermediate phase $FeF_2$ should first formed on the outside while Fe is located inside and may be screened or even isolated from the electrochemical system by the electrically and ionically insulating $FeF_2$ phase (Figure 4a inset, right). The correlation between phase distribution in an active particle and voltage hysteresis can be better seen in the schematic illustration in Figure 5. Even though the system is at the same state of lithiation, a $FeF_2$-rich (Li-poor) surface (during charge) and a Fe/LiF (Li-rich) surface (during discharge) will set the system at different potentials versus the $Li^+$/Li potential, which introduces a voltage gap (similar to concentration overpotential). This hysteresis caused by compositional inhomogeneity cannot be fully eliminated by voltage relaxation (zero-current) because it is very difficult, if not impossible, to make the relevant phases (or $Li^+$ distribution), such as $FeF_2$ ("Li-poor" phase) and Fe/LiF ("Li-rich" phase) at the states of lithiation close to $x_{Li} = 3$ per $FeF_3$, become spatially homogeneous simply by $Li^+$ diffusion. Such relaxation process also requires the migration of $F^-$ and $Fe^{2+}$ ions, which typically move very slowly. The situation is different from that of intercalation electrode materials (such as $LiCoO_2$ and graphite), in which during relaxation the $Li^+$ distribution can become homogenous more easily because there is no need for other ions or atoms to migrate. Here we use $V_{gap}$ to estimate the non-vanishing hysteresis at different states of lithiation, which is typically 300–500 mV for the NW electrode and 300–600 mV for the MW electrode when only $Fe^{2+}$ and $Fe^0$ are present (Figure 4d). When $Fe^{3+}$ is formed during charge and thus more significant inhomogeneity is introduced, the hysteresis gets significantly larger for both the NW (> 550 mV, $x_{Li}$ = 0.63) and the MW electrode (>700 mV, at $x_{Li}$ = 1.26 and 1.05). Now it is easier to understand why ~1 V voltage gap was observed even when the phases present are the same. For example, at states of lithiation $x_{Li}$ = ~1.8 (approximately the middle point in Figure 5), if we add the hysteresis caused by compositional inhomogeneity (~400 mV according to Figure 4d) to the hysteresis caused by reaction overpotential (300 mV during discharge + 70 mV during recharge = 370 mV according to ref. 20) and IR drop (~100 mV combined discharge and charge according to Figure 4a), we can expect a voltage gap ~1 V, which is consistent with what we observed during the galvanostatic discharge and charge experiments (Figure 5).

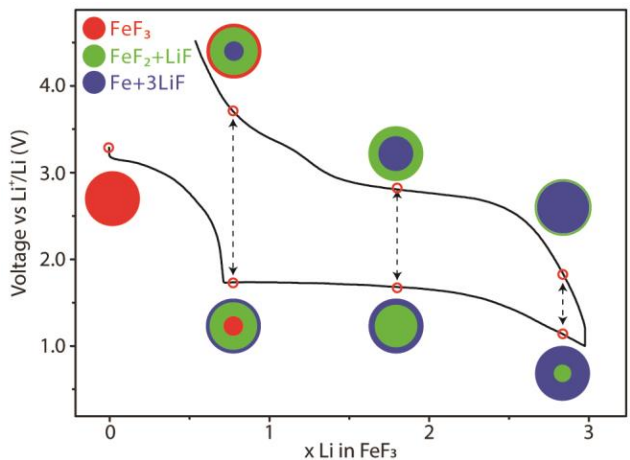

**Figure 5. Schematic illustration of the phase evolution in an active $FeF_3$ particle with compositional inhomogeneity and voltage hysteresis.** For the sake of clarity and simplicity, the volume change during the lithiation/delithiation process is ignored and compact layers are shown in this schematic illustration. Note that Ohmic voltage drop and reaction overpotential also contribute to the voltage hysteresis at non-zero current, in addition to that caused by compositional inhomogeneity. The voltage profile is collected from an $FeF_3$ NW electrode cycled at 1/10 C rate. The $FeF_3$ domain size is ~10–20 nm according to the TEM characterization.

This newly proposed compositional-inhomogeneity mechanism for driving hysteresis likely plays a role in many kinetically limited conversion materials where during Li insertion (extraction) the most reduced (oxidized) phases are present at the active surface and drive the potential down (up) compared the theoretical OCV. Interesting comparisons may also be made with Li-S and Li-$O_2$ batteries during discharging/charging: after voltage relaxation, a voltage gap remains in the Li-S system possibly due to presence of different $Li_2S_n$ (n = 8, 6, 4, 2, and 1) phases,[38] but approaches zero in the Li-oxygen system because there is only $Li_2O_2$.[39]

These new understandings suggest strategies to minimize the voltage hysteresis, which is important for improving the battery energy efficiency. One straightforward approach is constructing a composite electrode consisting of nanostructured active particles whose size must be comparable to the length scale of the conversion reaction (< 10 nm for $FeF_3$) and directly connected to electrically conductive scaffolds. A promising example could be embedding active materials between graphene layers to make a graphite intercalation compound (GIC).[40] This is expected to minimize the voltage hysteresis caused by compositional inhomogeneity as well as the iR drop. However, sufficient amount of active materials needs to be embedded so that the overall volumetric energy density is not severely compromised. Another approach that deserves further exploration is incorporating another cation or anion into the lattice (similar to the function of a "catalyst") to create a more disordered microstructure and improve ionic and electrical transport properties so that the reaction overpotential can be reduced. It is proposed that this approach can achieve higher energy density than the first one because no additional inactive components are introduced. As an example, the ternary fluoride $Cu_yFe_{1-y}F_2$ solid-solution exhibits smaller hysteresis than pure $FeF_2$ electrode[41]. One major challenge is to preserve the beneficial effect in repeated cycling, as Cu is rapidly lost through $Cu^+$ dissolution into the liquid electrolyte.[41, 42]

**CONCLUSION**

In summary, we elucidate the electrochemical reaction mechanism of the $FeF_3$ (and $FeF_2$) conversion electrode through integrated (*in situ*) experimental and theoretical studies. The phase evolution in the electrode is symmetrical during discharge and charge but the spatial distribution of the electrochemically active phases at the single-particle level, which is controlled by reaction kinetics, is very different. Such compositional inhomogeneity changes the way the active phases are connected to electrolyte ($Li^+$) and current collector (electron) during discharge and charge, which consequently introduces a voltage gap. This, along with reaction overpotential and the iR voltage drop, leads to the large voltage hysteresis. This understanding is contrary to the popular belief that attributes the voltage hysteresis of the $FeF_3$ electrode to asymmetric reaction pathways during discharge and charge. Further, since the issues that contribute to the hysteresis are kinetic in nature, it is hopeful that the voltage hysteresis can be reduced to a reasonable level (< 300 mV) by designing and optimizing material microstructure and electrode architecture. These results can help understand and minimize the voltage hysteresis in other conversion electrode materials, where compositional inhomogeneity was observed but not scrutinized,[42-44] despite their asymmetrical reaction pathways controlled by reaction kinetics. This work brings new hope to the development of high energy-density LIBs based on conversion chemistry and provide insights to the hysteresis problems in other next-generation battery chemistries, such as Li-S and Li-$O_2$.

**EXPERIMENTAL SECTION**

**Synthesis of FeF$_3$ samples.** FeF$_3$ NWs and MWs were synthesized by thermal dehydration of α-FeF$_3$·3H$_2$O NWs and MWs, respectively, at 350 °C for 2.5 h in argon atmosphere, based on previous work[18, 26]. Briefly, the precursor α-FeF$_3$·3H$_2$O NWs and MWs were first synthesized respectively by reacting different amounts of Fe(NO$_3$)$_3$·9H$_2$O and HF aqueous solution in ethanol at 60 °C for 18 h. The concentration ratio of $c(Fe^{3+})$ : $c(HF)$ : $c(H_2O)$ is 13.3 mM : 5560 mM : 6760 mM for the NW synthesis and 53.2 mM: 500 mM: 11575 mM for the MW synthesis. FeF$_2$ NWs were prepared by heating FeF$_3$ NWs (90 wt%) with a small amount of glucose (10 wt%) at 450 °C for 2.5 hour.

**Characterization**. Scanning electron microscopy (SEM) images were collected using a LEO 55 VP scanning electron microscope at 5 kV. Transmission electron microscopy (TEM) images and electron diffraction (ED) patterns were recorded using either a Philips FEI FM200 (200 kV) or a FEI Titan TEM (200 kV). Powder X-ray diffraction (PXRD) data were collected on a Bruker D8 diffractometer using Cu Kα radiation. Electrochemical measurements were performed on electrodes made of 70 wt% active material, 20 wt% carbon black and 10 wt% binder. The electrodes were packed into CR2032-type coin cells in an argon-filled glovebox, with Li metal as the counter/quasi-reference electrode, 1 M LiPF$_6$ in EC/DMC (1/1 by volume, BASF) as the electrolyte, and electrolyte-soaked polyethylene-polypropylene films as the separator. Galvanostatic cycling and galvanostatic intermittent titration technique (GITT) experiments were performed using either a Biologic SP-200 or a VMP-3 Potentiostat/Galvanostat controlled by EC-Lab software.

***In situ* X-ray Absorption Spectroscopy (XAS).** *In situ* X-ray absorption spectra were collected at beamline X18A, NSLS, BNL, using a perforated 2032-type coin cell with holes on both sides sealed by Kapton tapes. The electrodes were made of 70 wt% active material, 20 wt% carbon black and 10 wt% binder and coated on aluminum foil (25 μm thickness). The measurements were performed in transmission mode using a Si (111) double-crystal monochromator, which was detuned to ~35% of its original maximum intensity to eliminate the high order harmonics in the beam. A reference X-ray absorption spectrum of metallic Fe (*K*-edge 7112 eV) was simultaneously collected using a standard Fe foil. Energy calibration was done using the first inflection point of the Fe *K*-edge spectrum as the reference point. The X-ray absorption data were processed and analyzed using IFEFFIT-Athena, Artemis, and Atoms. Standard reference spectra from $FeF_3$, $FeF_2$, and Fe powders were collected to carry out spectrum fitting and estimate the ratio between different Fe oxidation states.

***In situ* TEM experiments.** *In-situ* STEM images, ED patterns were recorded at 200 kV in a JEOL2100F microscope. The *in situ* nano-battery consists of a copper half-grid (current collector), $FeF_3$ NWs supported on the amorphous carbon film (cathode) and Li metal (anode) was fabricated in an argon-filled glove box and transferred into the TEM chamber by using an argon-filled plastic bag. A thin passivation layer of $LiN_xO_y$ on the surface of the Li that formed due to brief exposure to air before transferring to the TEM chamber, acted as the solid electrolyte. The biasing probe was connected to the carbon membrane and the reaction was initiated by applying a negative bias typically at a value of 2 V.

**Computational methods.** All calculations were performed using Density Functional Theory (DFT) with the Vienna *ab initio s*imulation package (VASP)[45] and a plane wave basis set. The hybrid functional of Heyd, Scuseria and Ernzerhof (HSE06)[34] with Perdew-Burke-Ernzerhof (PBE)-type pseudopotentials[46] utilizing the projector augmented wave (PAW)[47] method was used for Fe, F and Li atoms. The valence electron configurations of Fe, F and Li atoms were Fe: $3p^6 3d^7 4s^1$, F: $2s^2 2p^5$, Li: $2s^1$. All calculations were performed with spin polarization enabled and with a plane wave cutoff energy set at least 30% larger than the maximum plane wave energy for the chosen set of pseudopotentials, equal to 520 eV. Reciprocal space integration in the Brillouin zone was performed with the Monkhorst-Pack scheme with k-point densities set for each material such that total energy convergence errors were less than 1 meV/cell[48].

Bulk Li, Fe, LiF, FeF$_2$ and FeF$_3$ materials were simulated within the $Im\bar{3}m$ (body centered cubic structure, Li and Fe), $Fm\bar{3}m$ (rocksalt structure), $P4_2/mnm$ (rutile structure) and $R\bar{3}c$ (rhombohedral structure) space groups, respectively. The lithiated FeF$_2$ structures, Li$_x$FeF$_2$ ($x$ = 0.25, 0.5, 1), and lithiated FeF$_3$ structures, Li$_x$FeF$_3$ ($x$ =0.25, 0.5, 1) were simulated as direct Li insertion into the interstitial sites of the rutile structure (Li$_x$FeF$_2$) and rhombohedral, monoclinic and defected trirutile structures (Li$_x$FeF$_3$). The monoclinic structure possesses the $Cc$ space group while the defected trirutile structure is based on the $P4_2/mnm$ space group and the ZnSb$_2$O$_6$ structure with the $2a$ Wyckoff sites vacant. These vacant sites serve as interstitial positions for direct Li insertion. The average voltage to insert Li $\bar{V}_{x_1 \to x_2}$ (in V/Li) from composition $x_1$ to composition $x_2$ in these structures is expressed as:

$$\bar{V}_{x_1 \to x_2} = -\frac{1}{(x_2 - x_1)} \left( E^{FeF_3/FeF_2}_{x_2} - E^{FeF_3/FeF_2}_{x_1} - (x_2 - x_1) E_{Li} \right) \quad (1)$$

where $E^{FeF_3/FeF_2}_{x_2}$ and $E^{FeF_3/FeF_2}_{x_1}$ are the calculated DFT energies of a lithiated FeF$_2$ or FeF$_3$ material with Li composition $x_2$ and $x_1$, respectively, and $E_{Li}$ is the DFT energy of metallic Li.

The phase stability of the Li-Fe-F system was analyzed by plotting the formation energies (relative to the pure elements Li, Fe, F) of each calculated compound at their respective compositions. The phase diagram is constructed by calculating the convex hull from these formation energies. Specific material compositions that are thermodynamically stable lie on the convex hull, while those that are unstable are above the convex hull.

## ASSOCIATED CONTENT

Supporting figures, tables, and movies, as noted in the main text. This material is available free of charge via the Internet at http://pubs.acs.org.

## AUTHOR INFORMATION

Correspondence and requests for materials should be addressed to D.M. (ddmorgan@wisc.edu) and S.J (jin@chem.wisc.edu).

**Notes**

The authors declare no competing financial interests.

## ACKNOWLEDGEMENTS

This research is supported by NSF grant DMR-1106184 and DMR-1508558 for the synthesis and structural characterization of the materials, and a UW-Madison WEI Seed Grant and Research Corporation SciaLog Award for the electrochemical studies. The *in situ* X-ray absorption spectroscopy experiments were performed at beamline X18A, National Synchrotron Light Source, Brookhaven National Laboratory, which are supported by the U.S. Department of Energy, Office of Basic Energy Sciences under Contract No. DE-AC02-98CH10886. R.J. and D.M. were supported by the NSF Software Infrastructure for Sustained Innovation (SI$^2$) award

No. 1148011. P.G. and F.W. were supported by the Laboratory Directed Research and Development (LDRD) program at Brookhaven National Laboratory. L.L also thanks Vilas Research Travel Awards for partially supporting the travel cost to the synchrotron facilities.
**REFERENCES**

1. Armand, M.; Tarascon, J. M. *Nature* **2008,** *451,* 652.

2. Bruce, P. G. *Solid State Ionics* **2008,** *179,* 752.

3. Goodenough, J. B.; Kim, Y. *Chem. Mater.* **2009,** *22,* 587.

4. Cabana, J.; Monconduit, L.; Larcher, D.; Palacín, M. R. *Adv. Mater.* **2010,** *22,* E170.

5. Bruce, P. G.; Freunberger, S. A.; Hardwick, L. J.; Tarascon, J.-M. *Nature Mater.* **2012,** *11,* 19.

6. Poizot, P.; Laruelle, S.; Grugeon, S.; Dupont, L.; Tarascon, J. M. *Nature* **2000,** *407,* 496.

7. Balaya, P.; Li, H.; Kienle, L.; Maier, J. *Adv. Funct. Mater.* **2003,** *13,* 621.

8. Morcrette, M.; Rozier, P.; Dupont, L.; Mugnier, E.; Sannier, L.; Galy, J.; Tarascon, J. M. *Nature Mater.* **2003,** *2,* 755.

9. Li, H.; Richter, G.; Maier, J. *Adv. Mater.* **2003,** *15,* 736.

10. Badway, F.; Cosandey, F.; Pereira, N.; Amatucci, G. G. *J. Electrochem. Soc.* **2003,** *150,* A1318.

11. Badway, F.; Pereira, N.; Cosandey, F.; Amatucci, G. G. *J. Electrochem. Soc.* **2003,** *150,* A1209.

12. Dreyer, W.; Jamnik, J.; Guhlke, C.; Huth, R.; Moskon, J.; Gaberscek, M. *Nature Mater.* **2010,** *9,* 448.

13. Cosandey, F.; Al-Sharab, J. F.; Badway, F.; Amatucci, G. G.; Stadelmann, P. *Microsc. Microanal.* **2007,** *13,* 87.

14. Doe, R. E.; Persson, K. A.; Meng, Y. S.; Ceder, G. *Chem. Mater.* **2008,** *20,* 5274.

15. Yamakawa, N.; Jiang, M.; Key, B.; Grey, C. P. *J. Am. Chem. Soc.* **2009,** *131,* 10525.

16. Wang, F.; Robert, R.; Chernova, N. A.; Pereira, N.; Omenya, F.; Badway, F.; Hua, X.; Ruotolo, M.; Zhang, R.; Wu, L.; Volkov, V.; Su, D.; Key, B.; Whittingham, M. S.; Grey, C. P.; Amatucci, G. G.; Zhu, Y.; Graetz, J. *J. Am. Chem. Soc.* **2011,** *133,* 18828.

17. Wang, F.; Yu, H.-C.; Chen, M.-H.; Wu, L.; Pereira, N.; Thornton, K.; Van der Ven, A.; Zhu, Y.; Amatucci, G. G.; Graetz, J. *Nature Commun.* **2012,** *3,* 1201.

18. Li, L.; Meng, F.; Jin, S. *Nano Lett.* **2012,** *12,* 6030.

19. Liu, P.; Vajo, J. J.; Wang, J. S.; Li, W.; Liu, J. *J. Phys. Chem. C* **2012,** *116,* 6467.

20. Ko, J. K.; Wiaderek, K. M.; Pereira, N.; Kinnibrugh, T. L.; Kim, J. R.; Chupas, P. J.; Chapman, K. W.; Amatucci, G. G. *ACS Appl. Mater. Interfaces* **2014,** *6,* 10858.

21. Parkinson, M. F.; Ko, J. K.; Halajko, A.; Sanghvi, S.; Amatucci, G. G. *Electrochim. Acta.* **2014,** *125,* 71.

22. Kim, S.-W.; Seo, D.-H.; Gwon, H.; Kim, J.; Kang, K. *Adv. Mater.* **2010,** *22,* 5260.

23. Li, C.; Mu, X.; van Aken, P. A.; Maier, J. *Adv. Energy Mater.* **2013,** *3,* 113.

24. Thorpe, R.; Rangan, S.; Whitcomb, R.; Basaran, A. C.; Saerbeck, T.; Schuller, I. K.; Bartynski, R. A. *Phys. Chem. Chem. Phys.* **2015**.

25. Zhang, W.; Duchesne, P. N.; Gong, Z.-L.; Wu, S.-Q.; Ma, L.; Jiang, Z.; Zhang, S.; Zhang, P.; Mi, J.-X.; Yang, Y. *J. Phys. Chem. C* **2013,** *117,* 11498.

26. Li, L.; Chen-Wiegart, Y.-c. K.; Wang, J.; Gao, P.; Ding, Q.; Yu, Y.-S.; Wang, F.; Cabana, J.; Wang, J.; Jin, S. *Nat. Commun.* **2015,** *6,* 6883.

27. Khatib, R.; Dalverny, A. L.; Saubanère, M.; Gaberscek, M.; Doublet, M. L. *J. Phys. Chem. C* **2013,** *117,* 837.

28. Boesenberg, U.; Marcus, M. A.; Shukla, A. K.; Yi, T.; McDermott, E.; Teh, P. F.; Srinivasan, M.; Moewes, A.; Cabana, J. *Sci. Rep.* **2014,** *4*.

29. Aykol, M.; Wolverton, C. *Phys. Rev. B* **2014,** *90,* 115105.

30. Laruelle, S.; Grugeon, S.; Poizot, P.; Dollé, M.; Dupont, L.; Tarascon, J.-M. *J. Electrochem. Soc.* **2002,** *149,* A627.

31. Hu, Y.-Y.; Liu, Z.; Nam, K.-W.; Borkiewicz, O. J.; Cheng, J.; Hua, X.; Dunstan, M. T.; Yu, X.; Wiaderek, K. M.; Du, L.-S.; Chapman, K. W.; Chupas, P. J.; Yang, X.-Q.; Grey, C. P. *Nature Mater.* **2013,** *12,* 1130.

32. Ponrouch, A.; Taberna, P.-L.; Simon, P.; Palacín, M. R. *Electrochim. Acta.* **2012,** *61,* 13.

33. Lowe, M. A.; Gao, J.; Abruna, H. D. *J. Mater. Chem. A* **2013,** *1,* 2094.

34. Heyd, J.; Scuseria, G. E.; Ernzerhof, M. *J. Chem. Phys.* **2003,** *118,* 8207.

35. Seo, D.-H.; Urban, A.; Ceder, G. *Phys. Rev. B* **2015,** *92,* 115118.

36. Chevrier, V. L.; Ong, S. P.; Armiento, R.; Chan, M. K. Y.; Ceder, G. *Phys. Rev. B* **2010,** *82,* 075122.

37. Weppner, W.; Huggins, R. A. *J. Electrochem. Soc.* **1977,** *124,* 1569.

38. Cuisinier, M.; Cabelguen, P. E.; Adams, B. D.; Garsuch, A.; Balasubramanian, M.; Nazar, L. F. *Energy Environ. Sci.* **2014,** *7,* 2697.

39. Cui, Z. H.; Guo, X. X.; Li, H. *Energy Environ. Sci.* **2015,** *8,* 182.

40. Wang, F.; Yi, J.; Wang, Y.; Wang, C.; Wang, J.; Xia, Y. *Adv. Energy Mater.* **2014,** *4,* 1300600.

41. Wang, F.; Kim, S.-W.; Seo, D.-H.; Kang, K.; Wang, L.; Su, D.; Vajo, J. J.; Wang, J.; Graetz, J. *Nat. Commun.* **2015,** *6,* 6668.

42. Hua, X.; Robert, R.; Du, L.-S.; Wiaderek, K. M.; Leskes, M.; Chapman, K. W.; Chupas, P. J.; Grey, C. P. *J. Phys. Chem. C* **2014,** *118,* 15169.

43. Wiaderek, K. M.; Borkiewicz, O. J.; Castillo-Martínez, E.; Robert, R.; Pereira, N.; Amatucci, G. G.; Grey, C. P.; Chupas, P. J.; Chapman, K. W. *J. Am. Chem. Soc.* **2013,** *135,* 4070.

44. Lee, D. H.; Carroll, K. J.; Chapman, K. W.; Borkiewicz, O. J.; Calvin, S.; Fullerton, E. E.; Meng, Y. S. *Phys. Chem. Chem. Phys.* **2014,** *16,* 3095.

45. Kresse, G.; Furthmüller, J. *Phys. Rev. B* **1996,** *54,* 11169.

46. Perdew, J. P.; Burke, K.; Ernzerhof, M. *Phys. Rev. Lett.* **1996,** *77,* 3865.

47. Kresse, G.; Joubert, D. *Phys. Rev. B* **1999,** *59,* 1758.

48. Monkhorst, H. J.; Pack, J. D. *Phys. Rev. B* **1976,** *13,* 5188.

**TOC GRAPHIC**

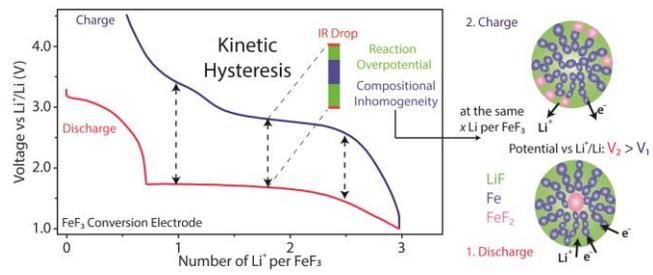